# AN ULTRACOOL STAR'S CANDIDATE PLANET

Steven H. Pravdo

Jet Propulsion Laboratory, California Institute of Technology

306-431, 4800 Oak Grove Drive, Pasadena, CA 91109; spravdo@jpl.nasa.gov

&

Stuart B. Shaklan

Jet Propulsion Laboratory, California Institute of Technology

301-451, 4800 Oak Grove Drive, Pasadena, CA 91109, stuart.shaklan@jpl.nasa.gov

## ABSTRACT

We report here the discovery of the first planet around an ultracool dwarf star. It is also the first extrasolar giant planet (EGP) astrometrically discovered around a main-sequence star. The statistical significance of the detection is shown in two ways. First, there is a 2 x $10^{-8}$ probability that the astrometric motion fits a parallax-and-proper-motion-only model. Second, periodogram analysis shows a false alarm probability of 3 x $10^{-5}$ that the discovered period is randomly generated. The planetary mass is $M_2$ = 6.4 (+2.6,-3.1) Jupiter-masses ($M_J$), and the orbital period is P = 0.744 (+0.013,-0.008) yr in the most likely model. In less likely models, companion masses that are higher than the 13 $M_J$ planetary mass limit are ruled out by past radial velocity measurements unless the system radial velocity is more than twice the current upper limits and the near-periastron orbital phase was never observed. This new planetary system is remarkable, in part, because its star, VB 10, is near the lower mass limit for a star. Our astrometric observations provide a dynamical mass measurement and will in time allow us to confront the theoretical models of formation and evolution of such systems and their members. We thus add to the diversity of planetary systems and to the small number of known M-dwarf planets. Planets such as VB 10b could be the most numerous type of planets because M stars comprise >70% of all stars. To date they have remained hidden since the dominant radial-velocity (RV) planet-discovery technique is relatively insensitive to these dim, red systems.

## 1. INTRODUCTION

Extrasolar planets have been found around stars where they were not expected (Marcy et al. 1998) and around stars where they were expected, but in unexpected places (Mayor & Queloz 1995). A lingering question of exoplanet discovery is whether the frequency of planetary systems around low-mass stars is similar to that established by radial velocity (RV) observations for solar-mass stars (Marcy et al. 2005a). Butler et al. (2006) argue that for their sample of 147 late K and M dwarfs ($0.2 - 0.6\ M_\odot$) planetary systems are ~3 times less common. This does not address stars with types later than about

M4. In this work we present the results of an astrometric search for planets around one such star.

## *1.1 VB 10*

VB 10 (= GJ 752B = V1298 Aql, van Biesbroeck 1944) is an astrometrist's dream in a nightmarish setting. It is nearby at ~6 pc, with low mass, and its background field is replete with astrometric reference stars. These properties can result in the measurement of a companion with a large signal-to-noise. However, they also present challenges. Its propinquity gives rise to high proper motion and frequent covering and, happily, uncovering of background stars. Its low mass results in a low luminosity making VB 10 relatively dim. The occasional interference of background stars and its dimness increase the astrometric noise. VB 10 is classified spectroscopically as M8 V (Kirkpatrick, Henry, & Simons 1996), an "ultracool" dwarf star with $T_{eff}$ ~2700 K (Schweitzer et al. 1996). Mass-luminosity relationships (MLRs) typically provide mass estimates for a star from its absolute *V, J, H, K* magnitudes but the VB 10 luminosity is below the lower limit of MLR applicability (Henry & McCarthy 1993, Henry et al. 1999, Delfosse et al. 2000). This places an upper limit on its mass of 0.08 $M_\odot$. The lower mass limit is 0.07 $M_\odot$ based upon its firm identification as a main-sequence star and the theoretical mass lower limit for such objects (Chabrier & Baraffe 2000). Hence the mass of VB 10 is already moderately constrained before our work begins. The age is ~1 Gyr (Martín, Basri, & Zapatero Osorio 1999) and it belongs kinematically to the old disk population of nearby dwarfs (Tinney & Reid 1998). The metallicity is believed to be near-solar, similar to its distant (~400 AU) proper-motion companion GJ 752A (Martín et al. 1999). A report of a planetary companion to VB 10 (Harrington, Kallarkal, & Dahn 1983) was later ruled out by subsequent observations (Monet et al. 1992, Harrington et al. 1993).

## *1.2 Other Observations of VB 10*

Radial velocity (RV) observations show no evidence for variations in VB 10 that might reveal the presence of a companion. At least 6 past measurements (Tinney & Reid 1998, Martín 1999, Martín et al. 2006, Basri & Reiners 2006) indicate a constant value for its heliocentric RV, with a mean of 35 km s⁻¹. However, these measurements have relatively low precisions of between 1.1 and 1.5 km s⁻¹, because VB 10 is a dim, red star. These precisions are more than 30 times worse than that achieved, for example, for the first M-star planetary system, GJ 876 (Marcy et al. 1998). Other techniques including direct imaging, astrometry, and speckle have failed to reveal a companion (Kirkpatrick, Henry, & Simons 1995).

## 2.   OBSERVATIONS AND RESULTS

### *2.1 Observations*

VB 10 was observed as part of the Stellar Planet Survey (STEPS) program (Pravdo et al. 2004) whose goal is to discover and characterize the low-mass companions—extrasolar giant planets (EGPs), brown dwarfs (BDs), and M-dwarfs--of a sample of previously-believed "single" M-dwarfs. The STEPS instrument is a CCD-camera mounted on the Cassegrain focus of the Palomar 200-inch (5-m) telescope and observes at 5500-7500 Å (closest to *R* band). Reflex motion of a target star around the system center-of-mass is compared to a grid of reference stars in the same field. The lowest mass companion (Pravdo, Shaklan, & Lloyd 2005) previously discovered with STEPS was GJ 802B, a ~60-$M_J$ BD.

Astrometric observations of VB 10 were taken over 9 years beginning with JD 2451438.64 = 17 September 1999. VB 10 is known to be an active star with occasional large flares (Linsky et al. 1995), variable Hα emission (Berger et al. 2008), and flaring X-ray emission (Fleming, Giampapa, & Schmitt 2000). However, there were no flares in our data and the intensity normalized to the reference frame varied by < ±6% despite non-photometric observing.

The reference frame contained 15 stars. Table 1 lists the reference stars and their USNO B1 designations (Monet 2003). The reference frame is used to make an affine transformation between observations (e.g. Eichhorn & Williams (1963), Shaklan et al. 1994). We tested for astrometric sensitivity to the choice of reference frame by using two different non-intersecting sets of reference stars, labeled "a" and "b" in Table 1 (column 1) and identified in Fig. 1. We found the same results described below for both independent reference frames, and for the frames combined. This is an important check to demonstrate that the signal does not arise from the reference frame. Figure 1 shows that the stars from both sets of reference frames surround VB 10 on all sides to ensure a non-biased transformation.

We performed analysis of the astrometric data as described by Pravdo et al. (2004). The only difference from that method is that we used the revised centroid determination method described in the following paragraph. As in our previous work, we calibrate for differential chromatic refraction (DCR) and minimize DCR effects by timing our observations to catch the target within 1.5 hours of the meridian. We avoid observing at dusk, dawn, and in moonlight, all of which can cause background intensity gradients.

The epochs and observing times for VB 10 are shown in columns 1 and 2 of Table 2. There are 21 observations over 11 epochs. Columns 6 and 7 show the relative RA and Decl. positions for VB 10. Columns 8 and 9 show the errors used for each data point. These are calculated from the standard error of the mean (SEM) added in quadrature with a 1 *mas* systematic error that is characteristic of STEPS (e.g., Pravdo et al. 2004). The SEMs were computed from the measured relative astrometric position of VB10 in the individual frames each night. They consist of the Poisson and atmospheric errors. Typically, 15-30 frames, each with 60-90 s integration times, were observed each night.

VB10 has high proper motion across a crowded field (Fig. 2). During the course of our observations, the wings of two background stars contaminated the frame close to VB 10, while a third passed almost directly behind it. On a frame-by-frame basis, we fitted a point spread function (PSF) model to these background stars and subtracted the fitted models from the frame. We were able to precisely center the models on the background stars by determining their relative positions in the frame at epochs when VB 10 was not in close proximity. The positions of VB 10 and the reference stars were then determined by centroiding their PSF cores. We performed this operation on the pixels whose flux values were ≥ 0.5 of the peak flux per star per image. This was done to minimize residual contamination from the wings of the PSF-subtracted background stars. We tested the efficacy of this approach by varying the minimum flux threshold to ensure that there was no systematic centroid shift caused by background stars. For the star that passed directly behind VB10 (see middle panel of Fig. 2) we were unsuccessful in reducing the centroid contamination at the 1 *mas* level and we have not included that data in our analysis. This accounts for a ~700 day gap in our data between epochs 2 and 3. This centroiding approach sacrifices some precision by not measuring the flux in the wings of the PSF, but it reduces background contamination to a negligible effect. It is important to note that for any uncorrected background objects passing through the wings of the VB10 PSF, the trajectory of the VB10 centroid motion is different on the RA and Decl. axes and is also aperiodic. The combined Poisson and systematic noise (columns 8 and 9) averages 1.7 *mas* for each point with a minimum of 1.2 *mas* and a maximum of 2.6 *mas* per point.

### 2.2 Parallax and Proper Motion Model

The motion of the star is very evident in our data (e.g., Fig. 2). We first fit these data with a parallax and proper motion (PPM) model using the USNO NOVAS code[1] to calculate the model. This gave a poor fit to the data with a chi squared, $X^2 = 94$ for 39 degrees of freedom (dof) or, $X^2_{dof} = 2.8$. The $X^2$ distribution probability that this model describes the data is very low, $2 \times 10^{-8}$. Fig. 3a show the residuals of these data, with errors from Table 2, folded over a 0.74-y period in anticipation of the results that follow. This indicates that there is an additional component of motion in the data.

### 2.3 Control Stars

We observed more than 30 stars in the STEPS program and for this data reduction we used three as control stars. The controls are within 2.5 hours in RA of VB 10 and are 4-36 times brighter in *R* (Table 3). When we observe stars we set the observation times of a single frame such that the CCD pixel full-well is ~2/3 full for the brightest useful star whether it is the target or a reference star. Thus despite the fact that VB 10 is dimmer than the control stars the integration times were still 60-90 seconds for all the fields and the Poisson limit was attained for all the targets. Fig. 3b-d shows the same analysis for the motion of these stars as illustrated in Fig. 3a for VB 10. Errors were calculated as before with the SEMs added in quadrature with a 1 *mas* systematic error. Table 3 also shows that 2 of the three have marginally acceptable fits to the PPM model, $X^2_{dof} = 1.2$ and 1.3, and the third has an unacceptable fit but still substantially better than VB 10. The last two columns in Table 3 compare literature values of PPM with those we determined. The STEPS confidence ranges are only estimated when there is not a good fit. There is a good correspondence with all the values except for G 212-57 which shows a large variation in its PM values compiled in SIMBAD. There may be non-PPM motion in G 212-57 that has been mistaken for PPM in some measurements. In any case none of the controls show a systematic motion as does the folded VB 10 data.

Table 2 (columns 3-5) also shows the epochs of observations of the control stars. As can be seen, there is complete overlap in these epochs with those of VB 10 although there are some differences on a nightly basis.

### 2.4 Periodogram Analysis of the VB 10 Residuals

We performed periodogram analysis of VB 10 and the control stars using a version of the Lomb-Scargle method (Scargle 1982). The periodogram measures the power at a range of periods that may be present in the data set. If a period is found at a given power level with the periodogram, the question then arises as to the significance of that period. Typically researchers answer that question by calculating the "false alarm probability" (FAP) of any period in their data. We do this by using a Monte Carlo program to create 100,000 random permutations of the VB 10 RA and Decl. data at the STEPS observing times. The FAP is taken to be the number of times any period exceeds a given power level divided by the total number of random permutations. Marcy et al. (2005b) state that "a 1% FAP [is] the criterion for candidate planets." Furthermore, Catanzarite et al. (2006) add that "[t]he detection of a signal exceeding the threshold corresponding to a 1% FAP is said to be at a 99% significance level."

---

[1] http://aa.usno.navy.mil/software/novas/novas_info.php

Following Catanzarite et al. (2006), we compute the periodograms in RA and Decl. separately, and then combine their power to form a single periodogram. This is a powerful advantage of astrometry whenever the astrometric signal appears in both axes with the same periodicity, as is the case here. While the individual axes may show a smaller FAP of, e.g. 1%, the combined periodogram will have an FAP of roughly the product, 0.01%. As long as the two axes have uncorrelated noise, the combined periodogram is a measure of the significance level of the combined signal.

Because our data are not equally spaced there is no definitive choice for the lower period limit. The minimum temporal spacing between observations of one day corresponds to a period of 0.0055 y and the minimum temporal spacing between epochs of 58 days corresponds to a period of 0.32 y. A comprehensive period search using our Keplerian model fitting code (§2.6) found no acceptable fits with periods < 0.43 y. The physics also favors longer periods because the fixed astrometric signal requires larger companion masses for shorter periods which then result in larger RV signals than have been observed (§2.6). Scargle (1982) notes that limiting the number of periods searched on physical grounds is a valid and important way to improve the detection efficiency. However, recognizing that the periodogram search is complementary to the model fitting process, for completeness we compute the periodogram to a minimum period of 0.1 y.

The periodogram of the VB 10 data is shown in Fig. 4a. Only two periods have FAP values smaller than the 1% criterion. The smallest FAP, 3 x $10^{-3}$ %, occurs at a 0.74-y period. This period has an FAP that is < 1% in each of the two independent axes. The other FAP is 5 x $10^{-2}$ % at a 0.43-y period.

Fig 4a also shows the periodograms of the control stars. The periodic power at all periods has FAPs well above 1% in all of the control stars. In particular none show any significant power at 0.74-y.

The peaks besides 0.74-y in the VB 10 periodogram can be explained as beats between the ~1-y time sampling of the STEPS data and the 0.74-y period. The sum beat period between 0.744 and 1 year is $P_{sum} = 1/[1/1+1/0.744] = 0.43$ y and the difference is $P_{diff} = 1/[1/1-1/0.744] = 2.9$ y, both of which are peaks in Fig. 4a. Fig 4b also demonstrates the origins of the beat periods. We created a synthetic data set with a noiseless 0.74-y period at the STEPS time sampling and created the periodogram shown in black in Fig 4b. This periodogram has nearly every feature of the real data periodogram in Fig. 4a, including the beat periods described above with the observed relative power in the periods. We did the same analysis with the period containing the second lowest FAP, 0.43-y, and plot that result as the blue periodogram in Fig. 4b. This shows that if the underlying period in the STEPS data were 0.43-y, a periodogram would show most the power there, inconsistent with the actual data.

With a FAP of 3 x $10^{-3}$ %, the significance level of the 0.74-y period in the VB 10 data is 99.997% (Catanzarite et al. 2006). This is consistent with the low probability that the PPM model alone can account for the VB 10 astrometric motion. We calculated the periodogram of the residuals to the data after subtraction of the best-fit model described below. This periodogram showed no significant power at any period (Fig. 4c).

### 2.5 Interpretations for the 0.74-y Period

There have been no confirmed reports of long-term periodicities over sixty years of VB 10 observations. Periodicities due to starspot rotations or stellar pulsations must be considered, but they are ruled out as significant contributors to the astrometric signal. Starspots on the primary are expected to add significant noise to astrometric observations at the <0.001 mas level (Catanzarite, Law, & Shao 2008), far below the signal reported herein. Symmetric pulsations of the primary would not add astrometric noise and asymmetric pulsations would again be too small to significantly contribute to the observed signal since the stellar diameter is <0.1 *mas* (Reffert et al. 2005). The astrometric signatures of gravitational instabilities within circumstellar disks are also believed to be too small (Rice et al. 2003) to result in a signal at the observed level. Photometric variations might also yield noise, but the periodogram of the VB 10 intensity normalized to the reference frame shows no significant power at any period, and in particular, no significant power at 0.74-y. The only remaining candidate for this periodicity is reflex motion of the primary due to a companion.

### 2.6 Keplerian Model

We next fit the motion data with a model that includes PPM and a two-body Keplerian orbit. There are 11 free parameters, 3 PPM, 7 Keplerian, and the secondary-to-total mass fraction. We fit the extant RV data simultaneously with the STEPS data, by subtracting the mean heliocentric RV, 35 km $s^{-1}$ (see RV references), from each measurement to obtain the VB 10 system RV at each of the observation times using the reported precisions as the uncertainties. The parameter search routine combines the features of a grid search, a Monte Carlo, and the Levenberg-Marquardt (L-M) method to thoroughly examine the complex multi-dimensional $X^2$ space. The grid parameters we use are planetary mass, $M_2$, semi-major axis, $a$, and eccentricity, $e$. For each grid point there are 10 random starting positions of the 8 other parameters that are optimized with the L-M algorithm. We searched values of $M_2$ from < 1 $M_J$ to the primary mass, $a$ > 0.01 AU, and all eccentricities. With this model, the minimum $X^2_{dof} = 0.83$. The fits confirm that the best-fit period is 0.74-yr. This $X^2_{dof}$ is lower than 1 because the contribution of the RV data to the $X^2$ is low due to the large RV errors. With the STEPS data alone, $X^2_{dof} = 0.93$.

The joint one-sigma confidence intervals for parameters are determined from the models that fit the combined data with a $X^2$ within the interval $X^2_{minimum}$ + 12.7, the statistical criterion for multi-parameter fits with 11 free parameters (Lampton, Margon, & Bowyer 1976). Within this $X^2$ interval there are only two period clusters, one at the 0.74-y best-fit period and another at a 2.9-y period that we interpreted above (§2.4) as the difference beat period. In what follows we select the 0.74-y period as the most likely to be correct. Note however, that if the 2.9-y period were correct, since the astrometric signal goes up as period$^{2/3}$, the corresponding companion mass goes down for the same signal.

The simultaneous Keplerian and PPM fit slightly changes the PPM values obtained in the PPM-only model. The parallax error in this fit is 0.9 *mas*. We add a 2 ± 1 *mas* correction from relative to absolute parallax for average fields at this galactic latitude and apparent magnitude (van Altena, Lee, & Hoffleit 1995). The total parallax uncertainty is thus 1.4 *mas*.

Figure 5 shows the motion of VB 10 after subtraction of the PPM component of the combined model. Portions of 8 orbital cycles are sampled. No single orbital cycle (~9 months) is sampled completely due to the inherent limitations of ground-based observing. In the 8 orbital cycles we sample multiple peaks, troughs, ascending portions, and descending portions of the orbits in both RA and Decl.

In the top two plots of Figure 6 these same data are folded over orbital phase and superimposed on the best-fit Keplerian model. Figure 6 plots two orbital periods to show the continuity near phase zero. This plot differs slightly from Fig. 3a because of the change in the PPM values between the PPM-only and the Keplerian plus PPM fits. The top and middle plots demonstrate that the motion is seen in both coordinates. The bottom plot shows the RV data obtained by subtracting the heliocentric system velocity from the measurements (Table 4). These data are useful only as upper limits. The astrometrically-determined planetary system does not violate these upper limits.

Figure 7 shows the best-fit Keplerian orbit and the data. We show the data epochal averages to clarify the display. The best-fit parameters and confidence limits for the orbital model are listed in Table 5. Figure 8 shows a plot of planetary mass, $M_2$, versus eccentricity, $e$, for the acceptable fits within the 1-sigma confidence interval. The ~10,000 models with acceptable fits are displayed from more than 750,000 trials. The location of the best-fit model is shown with a diamond.

The eccentricity of the orbit is not constrained by these data because of the time sampling and signal-to-noise. The fits show the trend we have reported before in our data of increasing possible companion mass with increasing eccentricity (cf., Cumming 2004, Shen & Turner 2004, O'Toole et al. 2009 for RV data analysis). In Fig. 8 we distinguish the Keplerian models by their RV amplitudes, where RV = 0.028 $M_2$ sin$i$ [$a$ ($M_1$+$M_2$) (1-$e^2$)]$^{-1/2}$ km s$^{-1}$, where $i$ is the inclination angle and $M_1$ is the primary mass. For models with RV < 1.5 km s$^{-1}$, the RV variations over the orbital phase are consistent with the current RV upper limit, and the companion mass is $M_2$ = 6.4 (+2.6,-3.1) $M_J$. This mass upper limit is below the deuterium burning minimum mass (Baraffe et al. 2003) of ~13 $M_J$ making the companion, VB 10b, an EGP. While some higher mass models, all of which appear in Fig. 8 with RV > 1.5 km s$^{-1}$, are not formally ruled out by the fitting process, we argue in the following section that they are unlikely to be correct. If the orbital period were ~2.9 y as discussed above the $M_2$ limits are 2.2-7.5 $M_J$ within the same RV limits.

### 2.7 High RV and Highly Eccentric Models

The RV data for this object help to rule out models with either a high secondary mass or with high eccentricity. Even if the system RV is up to 3 km s$^{-1}$, twice as high as the currently measured upper limit of 1.5 km s$^{-1}$, the maximum companion mass is still ≤ 13 $M_J$. Models with RV > 3 km s$^{-1}$ are allowed by all the astrometric and RV data, but only if the high-RV (near-periastron) orbital phases were missed by all the past measurements. Figure 9 illustrates two of the high-RV models depicted in Figure 8. In the top model the periastron was missed because it occurs between the RV measurements. In the bottom model, the first RV measurement deviates by 3-sigma from the model but the overall $X^2$ is acceptable. The probability that 6 measurements randomly distributed in orbital phase missed the high-RV phases in a given orbital model is estimated as $(1-f)^6$ where $f$ is the fraction of the orbital phase for which the RV > 1.5 km s$^{-1}$. For example, all the models shown in Figure 8 with $M_2 > 17$ $M_J$, have $f \geq 0.3$ and hence a probability less than 12% of being correct based upon the randomness of the past RV observations alone. The RV observations also rule out the nearly equal mass solution to the astrometric data as do previous speckle observations at larger separations (Henry 1991).

Highly eccentric models are also unlikely based upon the eccentricity distribution for the >300 known planets.[2] Less than 1% of the known planets have eccentricity $\geq 0.8$. Extremely high eccentricities are possible but unlikely in binary stars as well as in planetary systems. However, even if this were a highly eccentric orbit, Fig. 8 shows that the companion would still be a planet unless the RV > 3 km s$^{-1}$.

## 3. DISCUSSION

How does the VB 10 system look to an observer on the Earth? The star and planet would look about the same size since both the stellar (Beuermann, Baraffe, & Hauschildt 1999) and planetary (Baraffe et al. 2003) radii are expected to be ~0.1 $R_{\odot}$. The planet would be considerably dimmer in the infrared, but may approach a $J$-band ratio of ~4 x 10$^{-4}$ to the primary for the highest allowed planetary mass because of its ~400K internal temperature (Baraffe et al. 2003). The separation on the sky is ~60 mas leaving open the possibility that a future high-performance ground-based coronagraph or a moderate-performance space-based coronagraph could image the planet. The system is a difficult target for RV observations, and since transits probably do not occur, the dynamics of VB 10 and similar systems are likely to be explored by astrometry alone for the near-term future. However, future RV observations, even with the existing precision, could further rule out the unlikely, high eccentricity models described above.

VB 10b is the first planet with a primary star later than M4 and a dynamically-determined mass. It joins eight other M-dwarfs that have planetary systems with $M_2 \sin i$ determined by RV observations, but is the only one without the inclination-angle ambiguity, except for GJ 876b for which Benedict et al. (2002) separately determined $M_2$ and $i$, also astrometrically. It is also the heaviest planet with an M dwarf primary. The next most massive is OGLE-2005-BLG-071Lb, $M_2 = 3.8 \pm 0.4$ $M_J$ (Dong et al. 2009). The VB 10b mass fraction is 4.4-10.3%, which is considerably higher than OGLE-2005-BLG-071Lb, = 0.8 %. The fact that VB 10 is part of proper motion pair with the larger star GJ 752A (type M2.5) suggests the availability of a significant amount of circumstellar material during its formation process.

What kind of a planetary system might exist in VB 10 in addition to VB 10b? The habitable zone within which terrestrial planets can exist moves from ~1 AU as in the solar system to $\leq 0.1$ AU (Kasting, Whitmore, & Reynolds 1993) for the VB 10 primary spectrum. Furthermore a zone of planet orbital stability exists at distances <0.4 of the

---

separation (Musielak et al. 2005) between the two components, *viz.*, < 0.14 AU. The existence of an overlap between the habitable and stable zones here is in contrast with most of the known "hot Jupiter" planetary systems (Raymond 2006).

## Table 1: VB 10 Reference Frame Stars

| Ref. Star | USNO B1 Name[a] | $R^a$ | USNO B1 rel. flux[a,b] | STEPS rel. flux[b] |
|---|---|---|---|---|
| 1a | 0951-0432265 | 15.74 | 0.64 | 0.68 |
| 2a | 0951-0432090 | 14.16 | 2.73 | 2.30 |
| 3a | 0951-0432161 | 16.73 | 0.26 | 0.24 |
| 4a | 0951-0432298 | 14.45 | 2.09 | 1.07 |
| 5a | 0951-0432013 | 15.80 | 0.60 | 0.40 |
| 6a | 0951-0432384 | 16.47 | 0.33 | 0.32 |
| 7a | 0951-0432061 | 15.79 | 0.61 | 0.30 |
| 8a | 0951-0432479 | 15.93 | 0.53 | 0.46 |
| 9b | 0951-0432247 | 15.25 | 1.00 | 0.98 |
| 10b | 0951-0432159 | 15.31 | 0.95 | 0.98 |
| 11b | 0951-0432172 | 16.31 | 0.38 | 0.25 |
| 12b | 0951-0432258 | 15.14 | 1.11 | 0.31 |
| 13b | 0951-0432029 | 15.36 | 0.90 | 0.81 |
| 14b | 0951-0432427 | 15.18 | 1.07 | 2.27 |
| 15b | 0951-0432082 | 16.00 | 0.50 | 0.20 |

[a]From USNO B1 catalog (Monet et al. 2003), [b]Flux relative to VB 10

**Table 2: Observations of VB 10 and Comparisons Stars**

| Epoch | Observing Times (JD + 2451438.64) | | | | VB 10 Relative Position (*mas*) | | VB10 Error (*mas*) | |
|---|---|---|---|---|---|---|---|---|
| | VB 10 | GJ 777B | GJ 1253 | G 212-57 | RA | Decl | RA | Decl. |
| 1 | 0.00 | 0.05 | 0.07 | 0.13 | 0.0 | 0.0 | 1.6 | 1.4 |
| | | 1.05 | 1.07 | 1.11 | | | | |
| 2 | 294.15 | 294.21 | 294.23 | | -312.9 | -1030.2 | 1.8 | 1.8 |
| | | 295.21 | 295.24 | 295.31 | | | | |
| | 296.15 | 296.20 | | 296.30 | -321.2 | -1040.2 | 1.8 | 1.6 |
| 3 | 1022.18 | 1022.24 | 1022.26 | 1022.31 | -1479.4 | -3734.9 | 1.6 | 1.8 |
| | 1023.18 | 1023.25 | 1023.27 | 1023.31 | -1479.6 | -3741.2 | 1.9 | 2.6 |
| 4 | 1385.17 | 1385.23 | 1385.26 | 1385.31 | -2061.0 | -5092.7 | 1.3 | 1.7 |
| 5 | 1449.03 | 1449.07 | 1449.10 | 1449.15 | -2319.2 | -5380.2 | 1.5 | 2.0 |
| | 1450.04 | 1450.12 | 1450.09 | 1450.17 | -2319.9 | -5386.7 | 1.8 | 1.6 |
| | 1451.05 | 1451.08 | 1451.10 | 1451.14 | -2322.7 | -5389.2 | 2.0 | 2.2 |
| 6 | 1760.15 | 1760.20 | 1760.23 | 1760.27 | -2688.3 | -6487.7 | 1.5 | 1.7 |
| | 1761.15 | 1761.20 | | | -2691.9 | -6493.0 | 1.8 | 1.6 |
| | | | 1762.23 | 1762.27 | | | | |
| 7 | 1823.02 | 1823.06 | 1823.08 | 1823.13 | -2931.9 | -6786.9 | 1.9 | 1.6 |
| | 1824.01 | 1824.06 | 1824.08 | 1824.13 | -2936.4 | -6785.8 | 1.6 | 2.4 |
| | 1825.01 | 1825.04 | 1825.06 | 1825.12 | -2940.4 | -6794.0 | 1.2 | 1.8 |
| 8 | | | 2140.17 | 2140.23 | | | | |
| | 2141.14 | 2141.17 | | | -3348.6 | -7920.5 | 1.6 | 1.5 |
| | 2142.13 | 2142.17 | | | -3356.0 | -7927.7 | 1.4 | 1.3 |
| 9 | 2480.18 | 2480.22 | 2480.25 | 2480.28 | -3823.7 | -9173.8 | 1.5 | 1.6 |
| 10 | 2538.05 | 2538.10 | 2538.12 | 2538.16 | -4063.4 | -9431.8 | 1.7 | 1.7 |
| | 2539.03 | 2539.07 | 2539.09 | 2539.16 | -4069.1 | -9436.3 | 1.6 | 1.5 |
| 11 | 3276.01 | 3276.06 | 3276.09 | 3276.14 | -5267.4 | -12195.4 | 1.9 | 2.1 |
| | 3277.03 | 3277.07 | 3277.09 | 3277.12 | -5271.6 | -12195.9 | 2.1 | 1.7 |

**Table 3: Control Stars' Comparison to VB 10**

| Star ID | $R$ mag | No. Obs. | PPM $X^2_{dof}$ | SIMBAD[3] PPM[a] | STEPS PPM[a] |
|---------|---------|----------|-----------------|------------------|--------------|
| GJ 777B | 13.0 | 23 | 1.3 | 56 ± 4<br>689<br>-515 | 63.0 (+1.1,-0.6)<br>684.6 (+0.1,-0.2)<br>-516.9 ± 0.1 |
| GJ 1253B | 14.0 | 21 | 1.2 | 106 ± 4<br>269<br>549 | 104.7 ± 0.9<br>259.1 ± 0.1<br>543.8 ± 0.1 |
| G212-57 | 11.7 | 21 | 1.6 | 59 ± 11<br>126<br>149 | 60.5 (+0.8,-0.7)<br>279.7 ± 0.1<br>0.1 (+0.1,-0.2) |
| VB 10 | 15.6 | 21 | 2.8 | 164.3 ± 3.5<br>-614<br>-1368 | 172.1 (+1.3,-0.7)<br>-588.6 (+0.1,-0.2)<br>-1360.7 ± 0.1 |

[a]**PPM are the parallax, RA, and Decl. proper motions in *mas*.**

**Table 4: RV Observations of VB 10**

| JD | RV (m sec⁻¹) | Ref. |
|----|--------------|------|
| 2448812.5 | -100 ± 1500 | Tinney & Reid 1998 |
| 2449681.5 | 300 ± 1400 | Basri & Reiners 2006 |
| 2449788.5 | 0 ± 1400 | Basri & Reiners 2006 |
| 2449874.5 | 0 ± 1100 | Martín 1999 |
| 2452075.5 | 0 ± 1500 | Martín et al. 2006 |
| 2452215.5 | 0 ± 1500 | Martín et al. 2006 |

**Table 5: STEPS VB 10 Astrometric Measurements and Literature Values**

| Quantity | Literature Value | STEPS Measurements |
|---|---|---|
| **RA (2000)** | 19h 16m 57.605s[a] | -- |
| **Decl. (2000)** | +05° 09' 01.61'' | -- |
| **Proper motion (*mas*)** | 1479.4 ± 15.3[b] | 1483.1 ± 0.3 |
| **Position angle (deg)** | 202.9 ± 0.5 | 203.40 ± 0.01 |
| **Parallax-absolute (*mas*)** | 170.26 ± 1.37[c] | 171.6 (+1.4,-1.3) |
| **Period (y)** | -- | 0.744 (+0.019,-0.008) |
| **Total mass ($M_\odot$)** | -- | 0.0841 (+0.0043,-0.0108) |
| **Primary mass, $M_1$ ($M_\odot$)** | -- | 0.0779[a] |
| **Companion mass, $M_2$ ($M_J$)** | -- | 6.4 (+2.6,-3.1) |
| **Semi-major Axis (AU)** | -- | 0.360 (+0.006,-0.016) |
| **Semi-major Axis (*mas*)** | -- | 61.8 (+1.0,-2.7) |
| **Eccentricity** | -- | <0.98 |
| **Inclination (deg)** | -- | 96.9 (+7.4,-1.8) |
| **Long. of ascending node (deg)** | -- | 38.7 (+4.8,-3.3) |
| **Arg. of periastron, epoch** | -- | Unconstrained |

[a]Lepine & Shara 2005, [b]Tinney 1996, [c]Gould & Chanamé 2004

## ACKNOWLEDGMENTS

The research described in this paper was performed in part by the Jet Propulsion Laboratory, California Institute of Technology, under contract with the National Aeronautics and Space Administration. We thank the staff at Palomar Observatory for their assistance. We thank J. Catanzarite for providing the periodogram code and for helpful discussions. This research has made use of the NASA/IPAC archive, SIMBAD, NASA ADS, 2MASS Survey, and the Digitized Sky Survey. The supercomputers used in this investigation were provided by funding from the JPL Office of the Chief Information Officer. Copyright 2009 California Institute of Technology. Government sponsorship acknowledged.

**FIGURE CAPTIONS**

1. STEPS image of the VB 10 (red star symbol) field that depicts the two separate sets of reference stars (green triangles and blue squares) that were used to determine the relative position of VB 10 over the observing period. The brightest star in the northwest of the image is GJ 752A.

2. This series of STEPS images show the passage of VB 10 over a background star (center picture).

3a-d. Residuals of the motion (RA above and Decl. below) after subtraction of the best-fit PPM models and phased over a 0.74-y period for (A) VB 10, (B) GJ 777B, (C) GJ 1253, and (D) G 212-57.

4a. The periodogram of the VB 10 motion data shows the 0.74-yr orbital period as the highest peak with the lowest FAP = 3 x $10^{-5}$, or a ~4-sigma detection. We also show the periodograms of three control targets that were observed at the same epochs in the same system configuration. These comparison targets do not show the 0.74-yr period nor any other significant period.

4b. A comparison periodogram with a model of a pure 0.74-yr period (black) and the same sampling as the VB 10 motion data. Note the nearly exact correspondance between the peaks in the synthetic data and the peaks in the VB 10 motion data demonstrating that they are beats between the 0.74-yr period and the time sampling. The periodogram of a pure 0.43-y period (blue) shows that the 0.74-y period would not be the most prominent period if the underlying period were 0.43-y.

4c. The periodogram of the residuals after subtraction of the best-fit model. No significant power is seen at any period.

5. The motion of VB 10 as a function of elasped time after subtraction of the PPM portion of the combined model. This shows the portions of the 8 orbital cycles that were sampled by the data.

6. The motion of VB 10 in RA (top) . Decl. (middle), and RV (bottom). The astrometric data and model are after subtraction of the best-fitting PPM portion of the combined model and after folding over the best-fitting period. All of the existing data spanning 11 orbital cycles are displayed. The data are repeated for two orbital phases. The curve shows a best-fit model with 0.744-y period and 6.4-$M_J$ companion.

7. The Keplerian orbit of VB10b with the average data from each epoch shown.

8. The VB 10b mass, $M_2$, is shown as a function of eccentricity, $e$, for the acceptable fits to the STEPS + RV data. Different colors show different RV intervals for the fits.

9. Two high RV models that fit the VB 10 astrometric and RV data but are unlikely because either the RV observations had to all miss the periastron (top) or a poor fit to some RV data points is compensated by the astrometric data points (bottom).

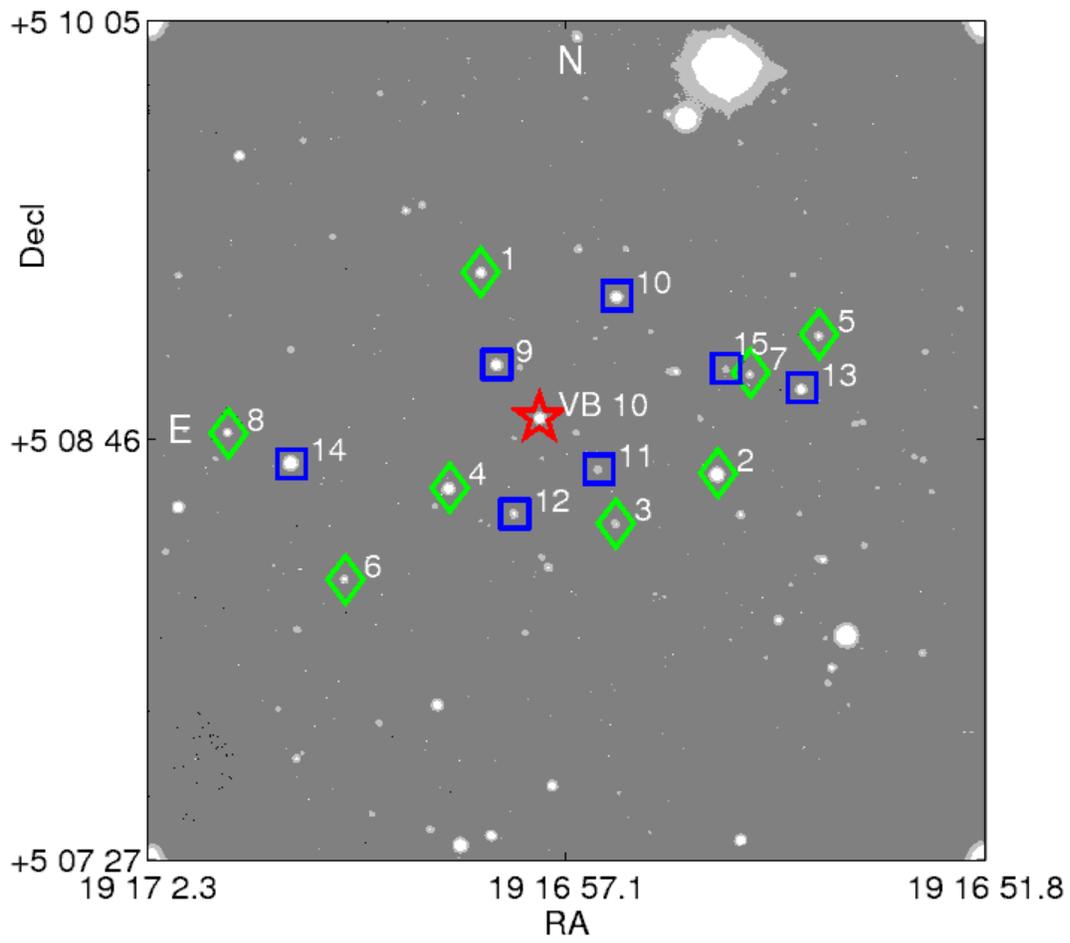

**Figure 1**

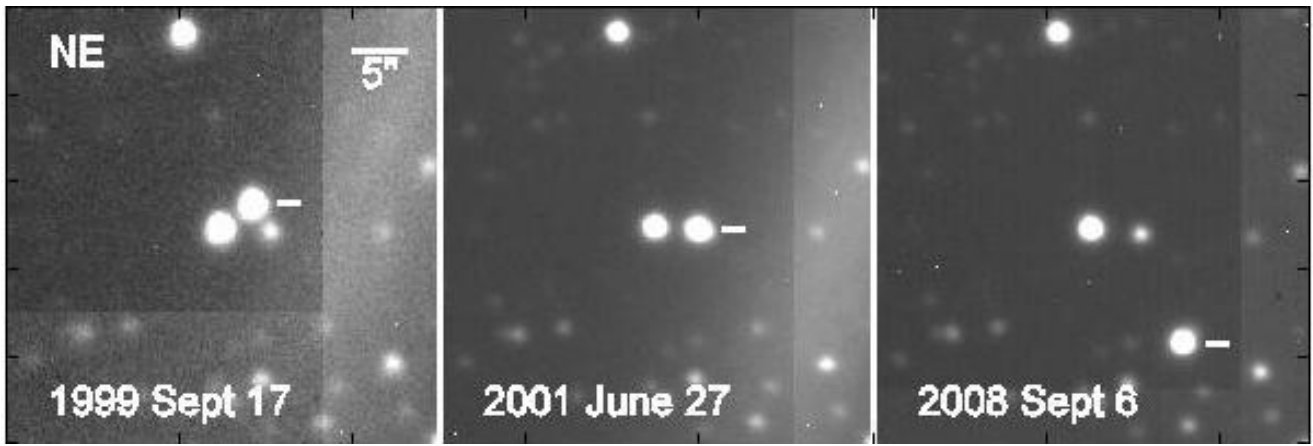

**Figure 2**

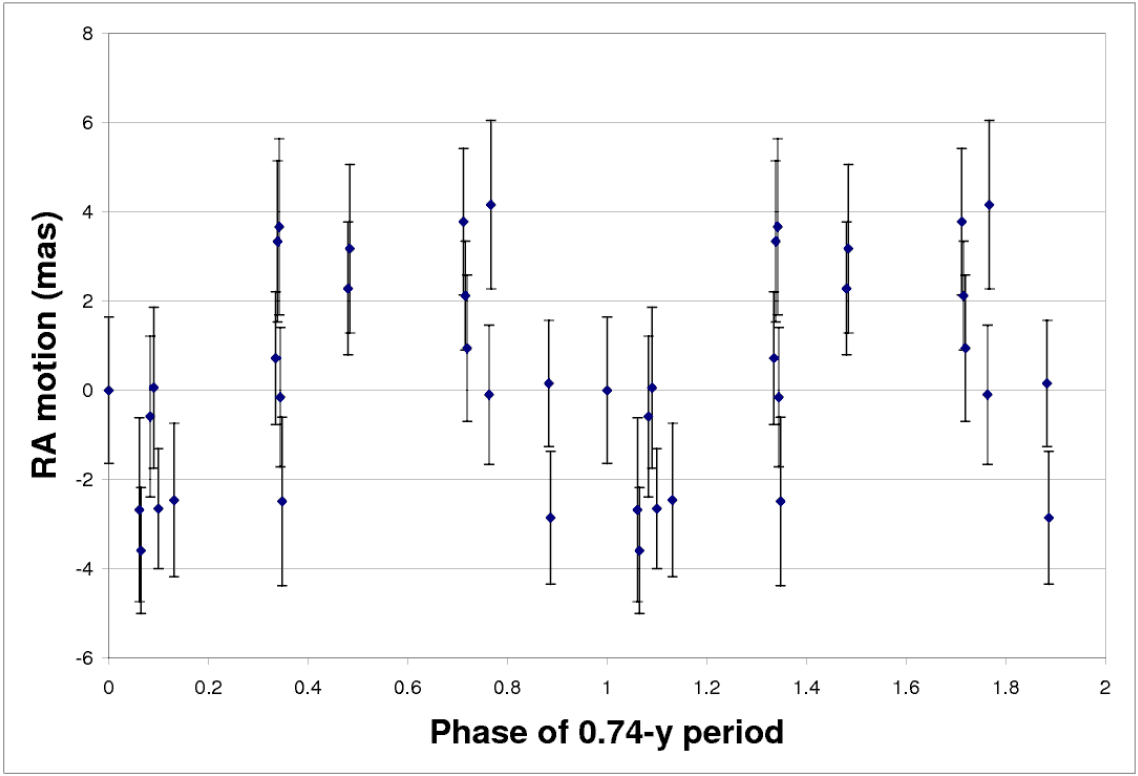

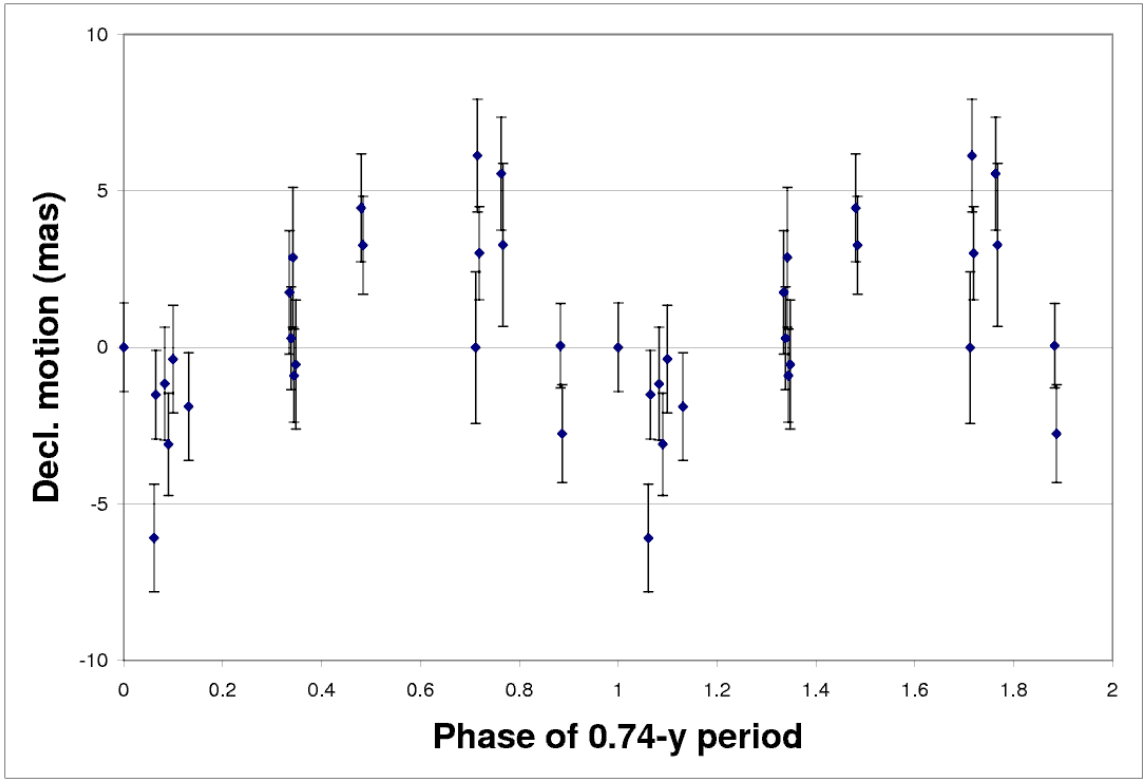

Figure 3a

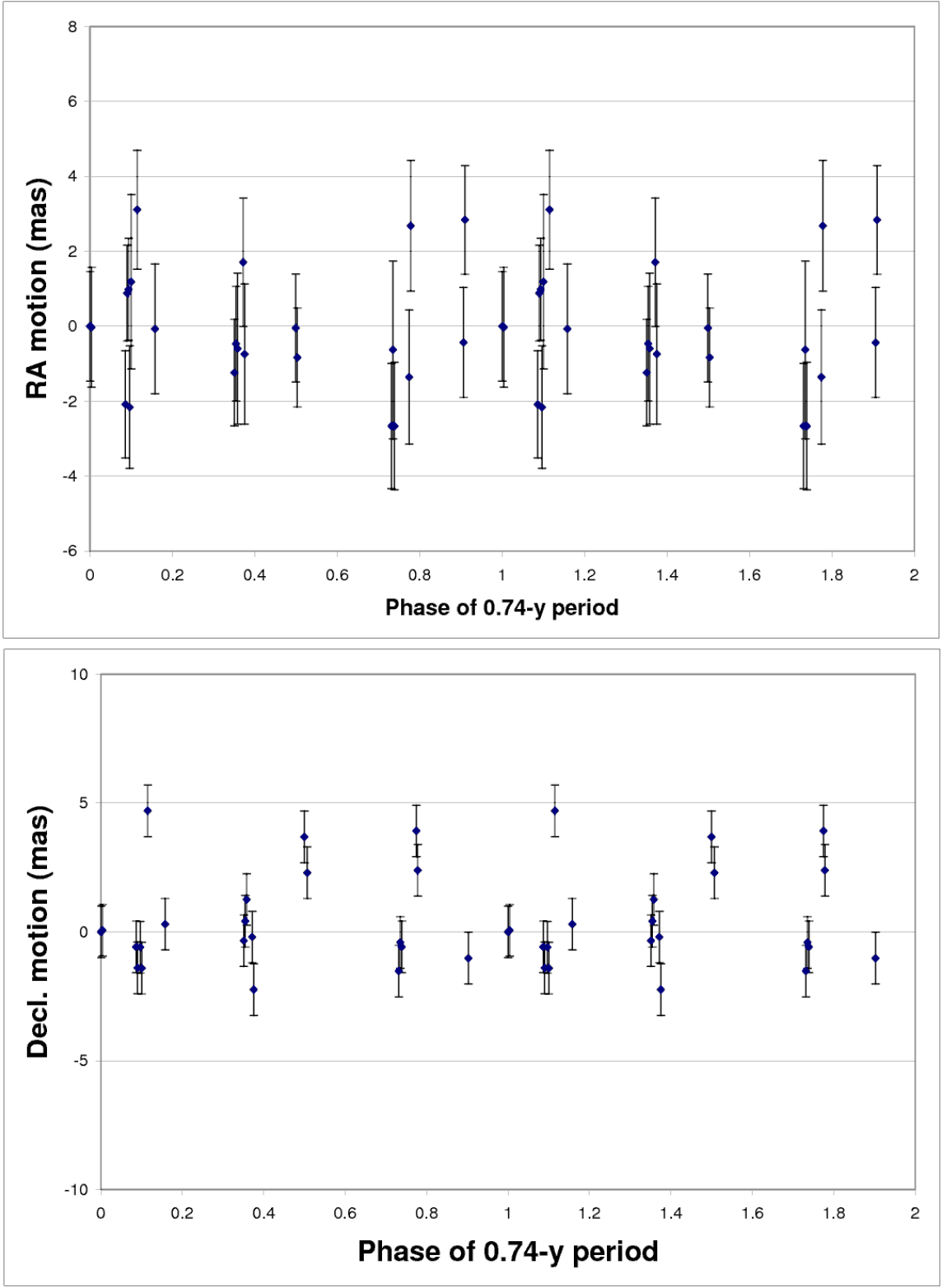

**Figure 3b**

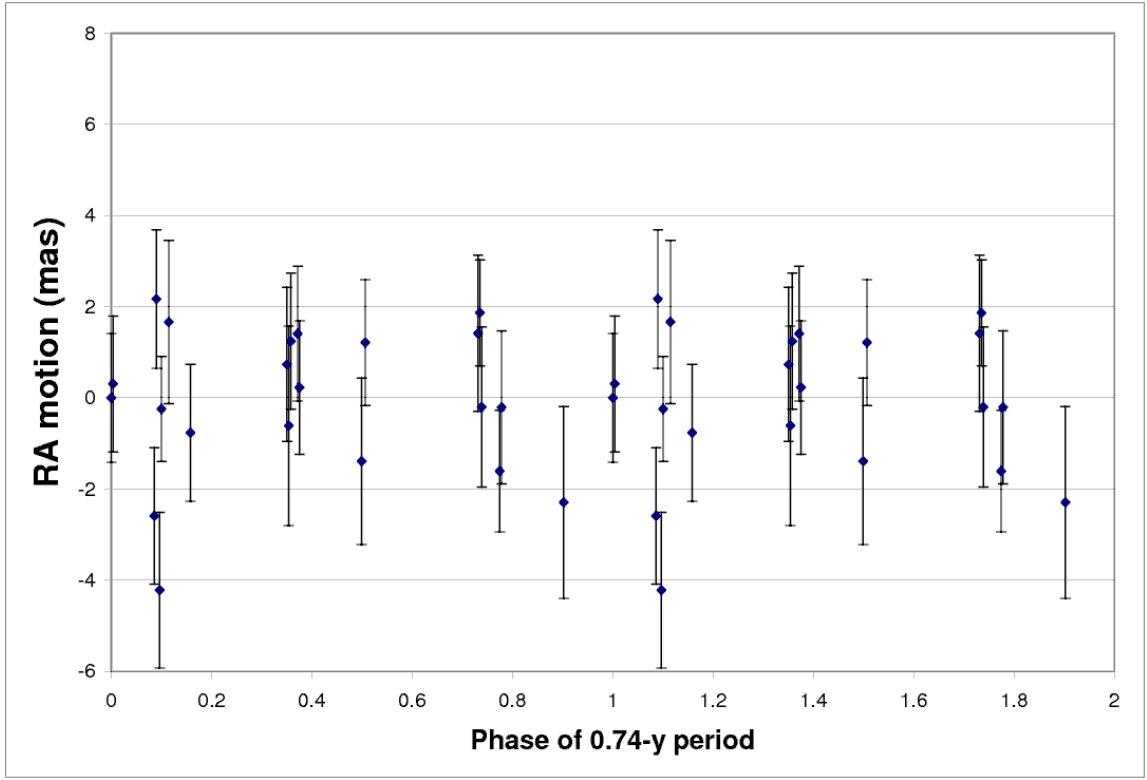

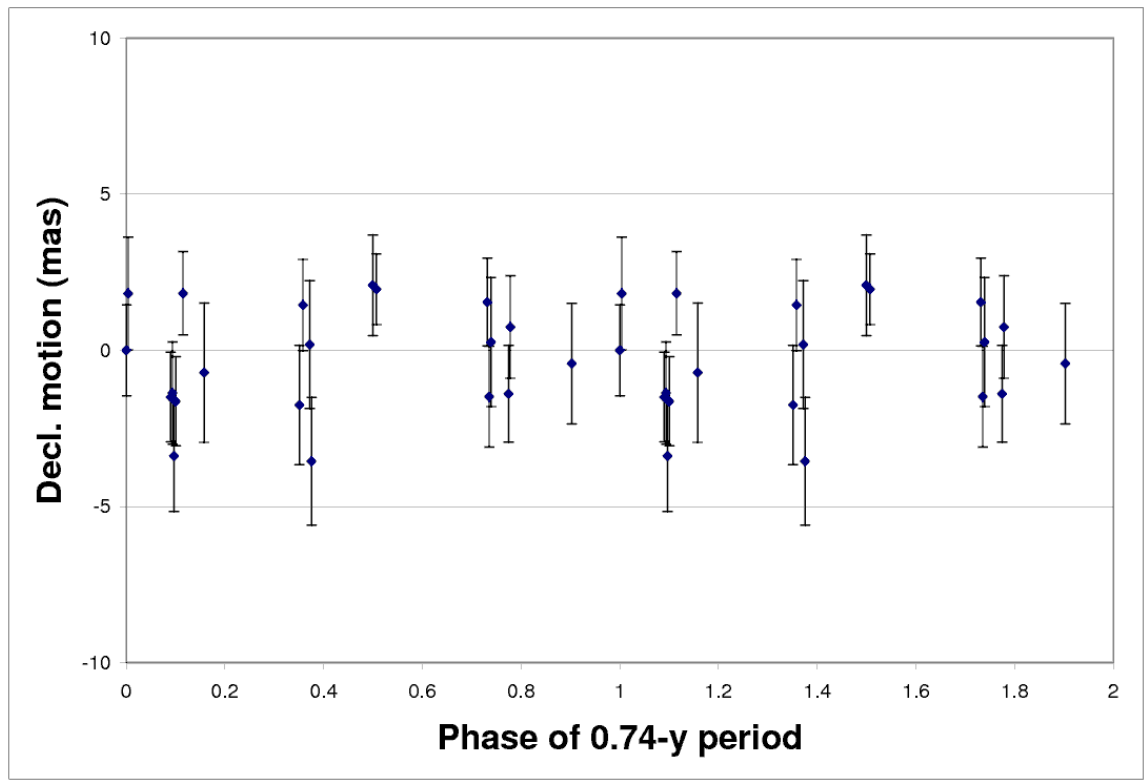

**Figure 3c**

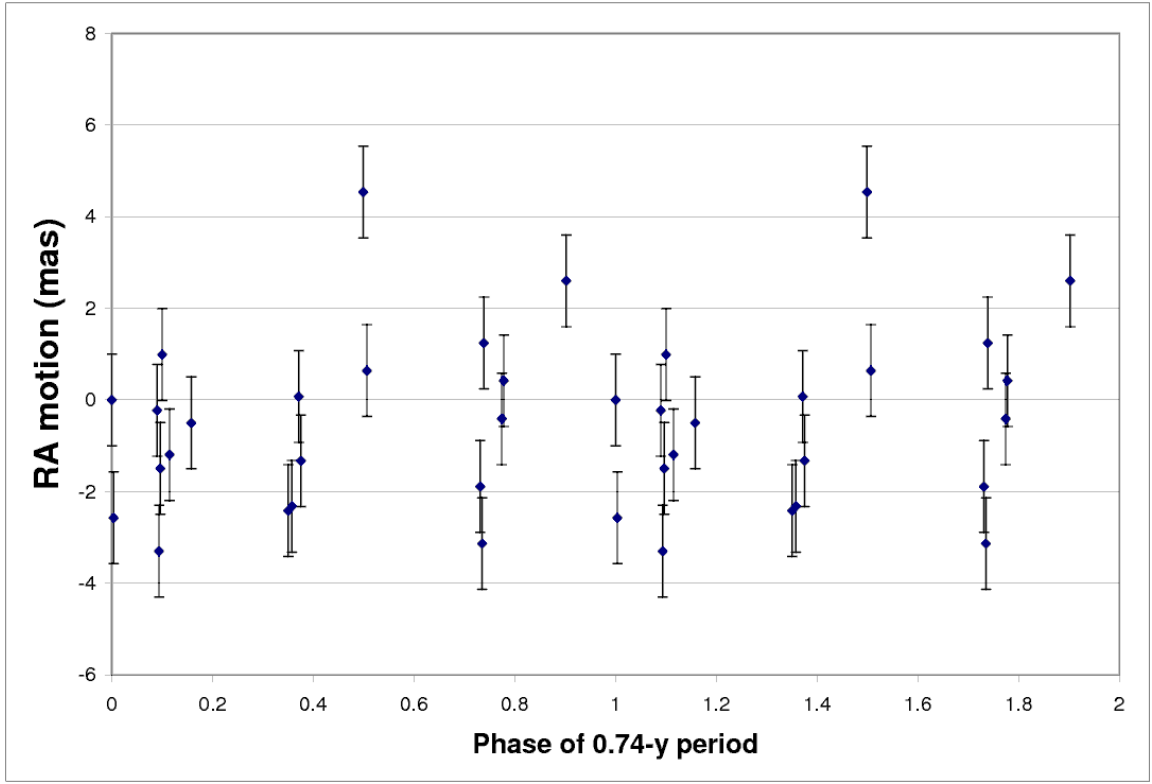

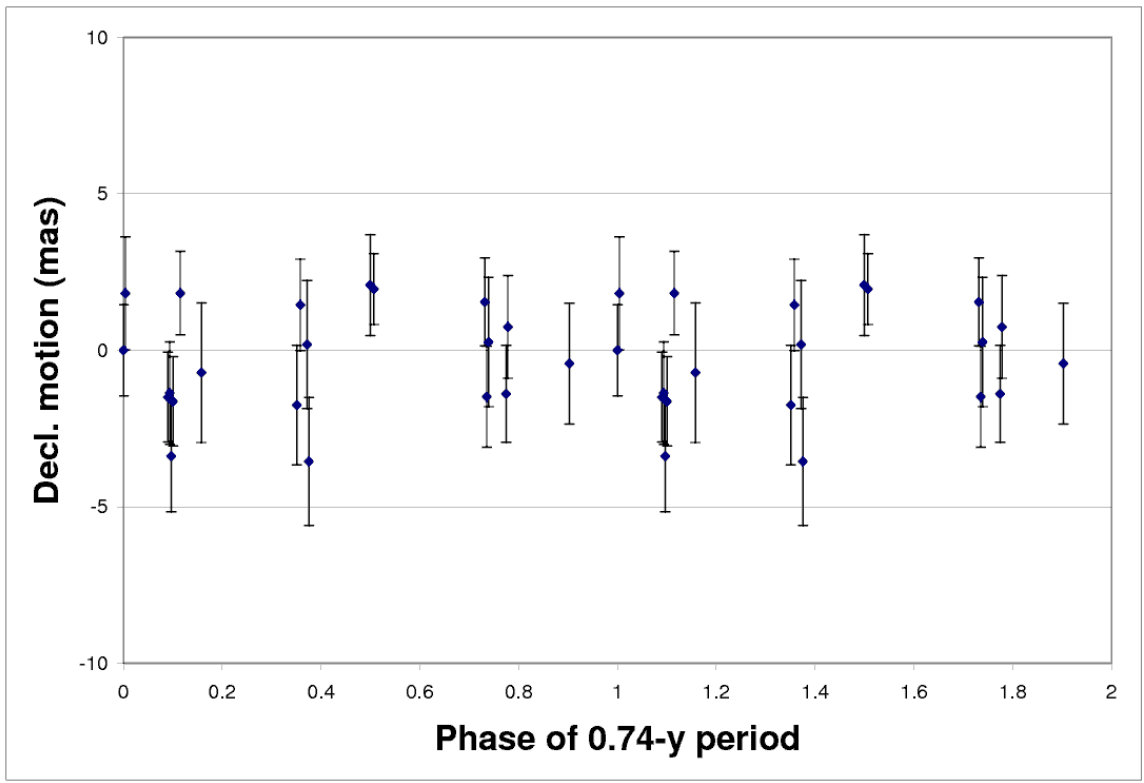

**Figure 3d**

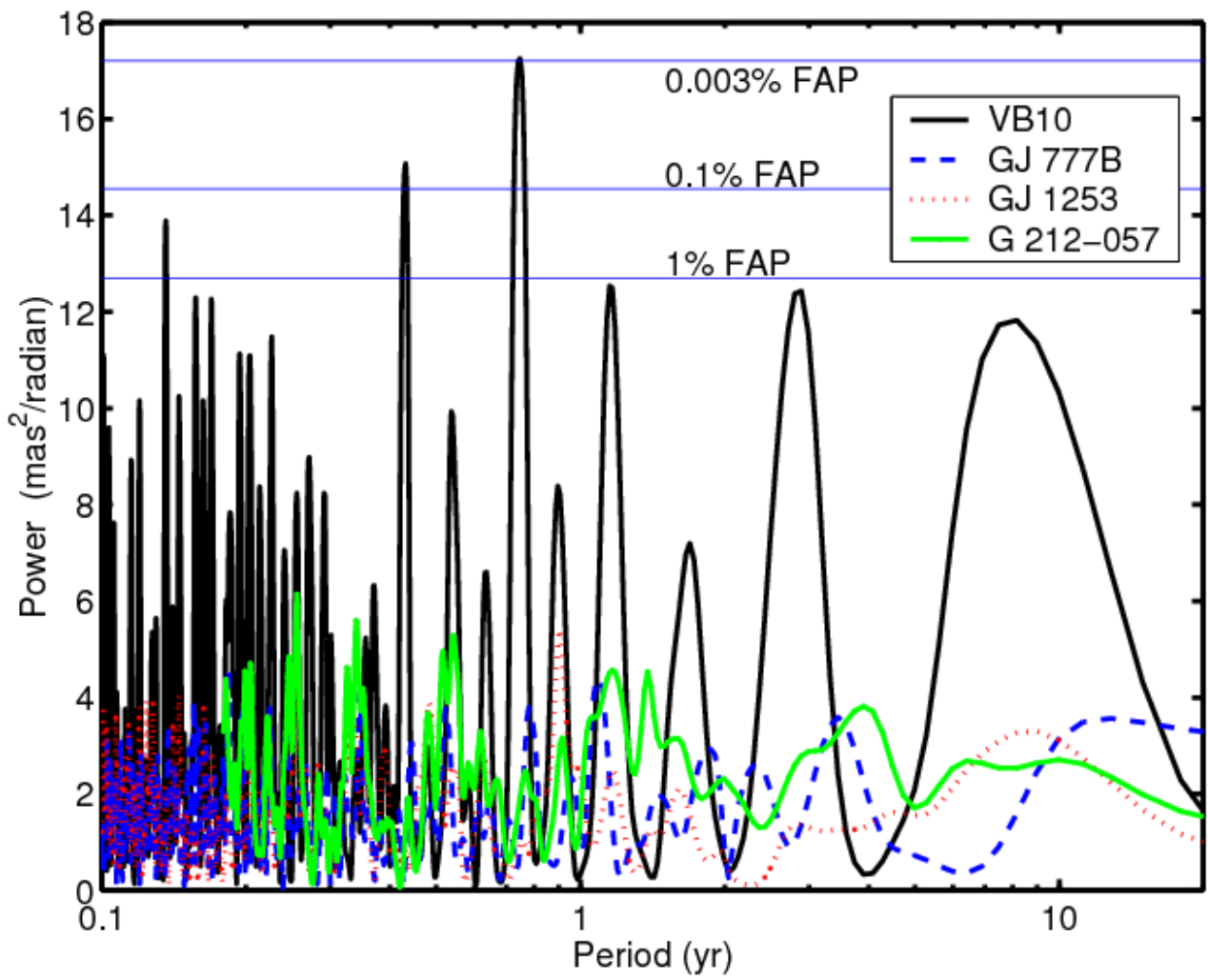

**Figure 4a**

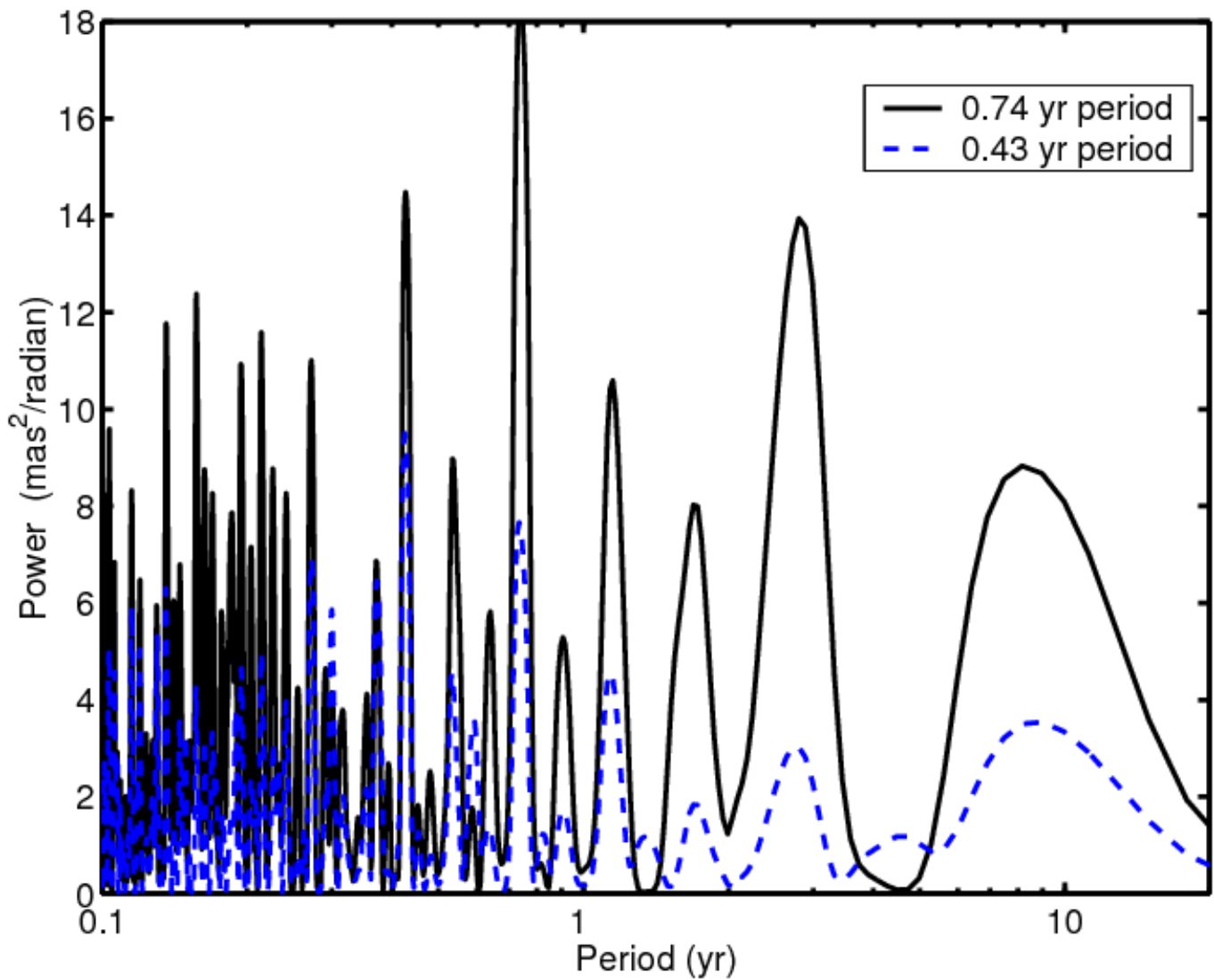

**Figure 4b**

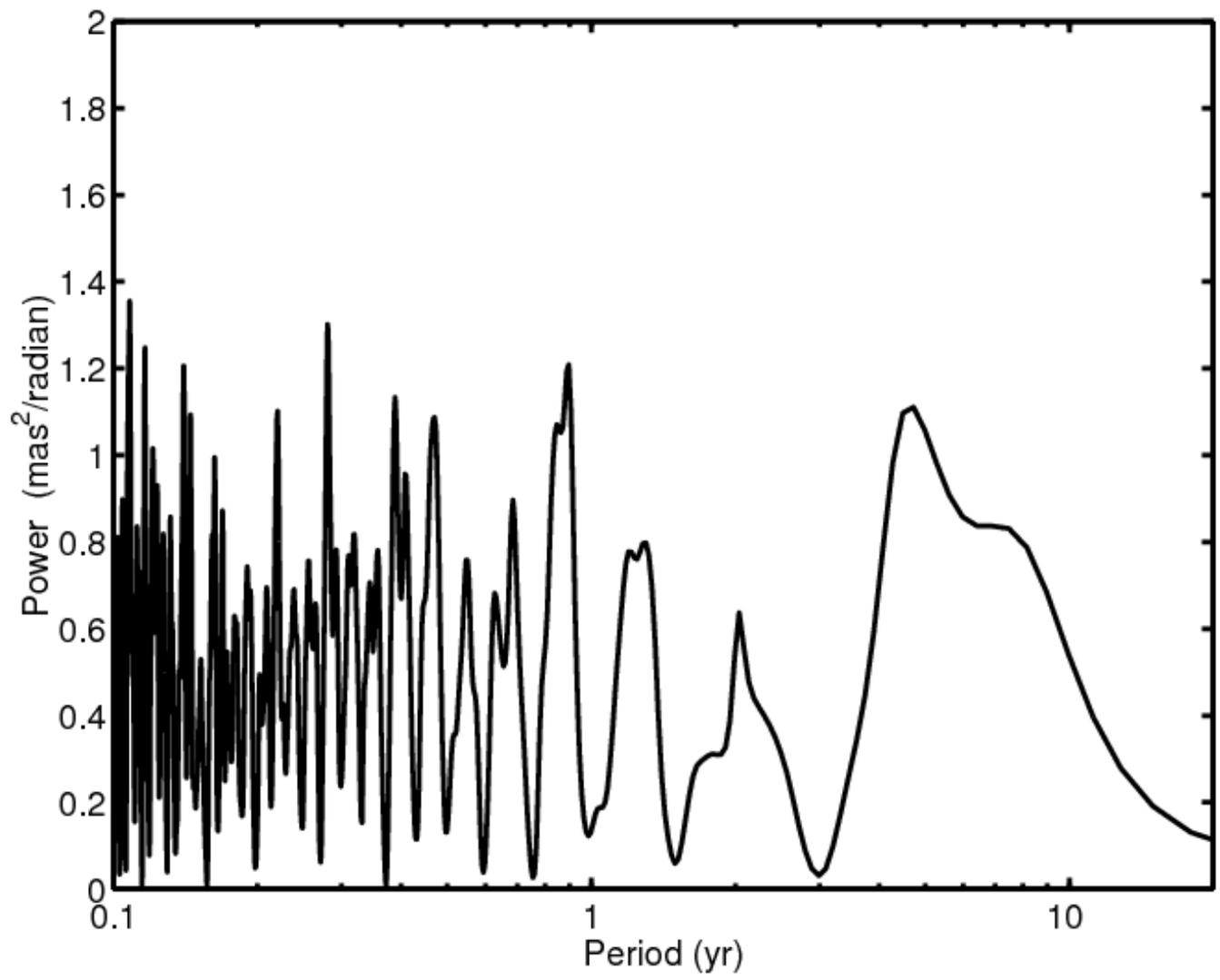

**Figure 4c**

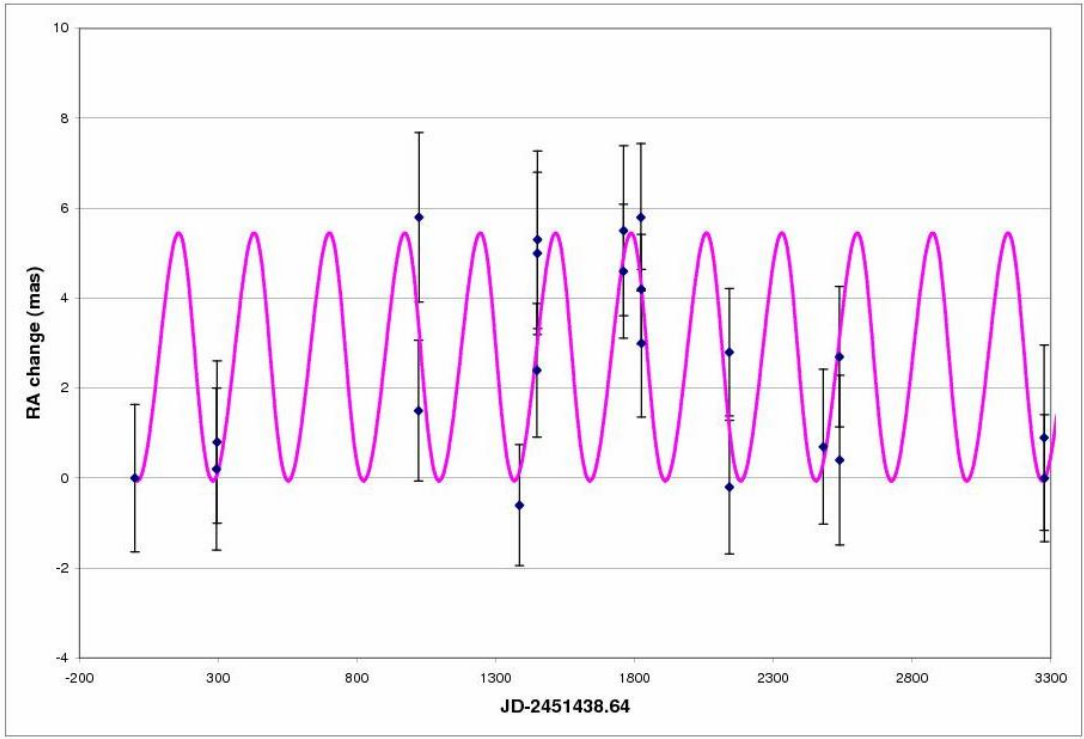

**Figure 5a**

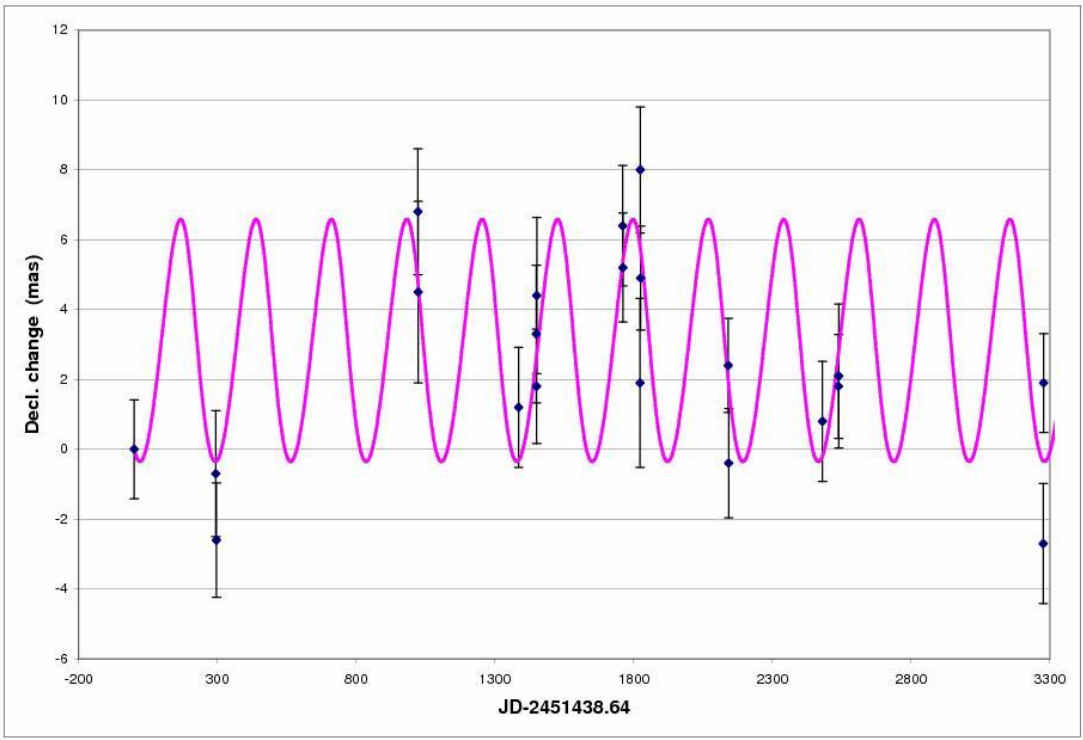

**Figure 5b**

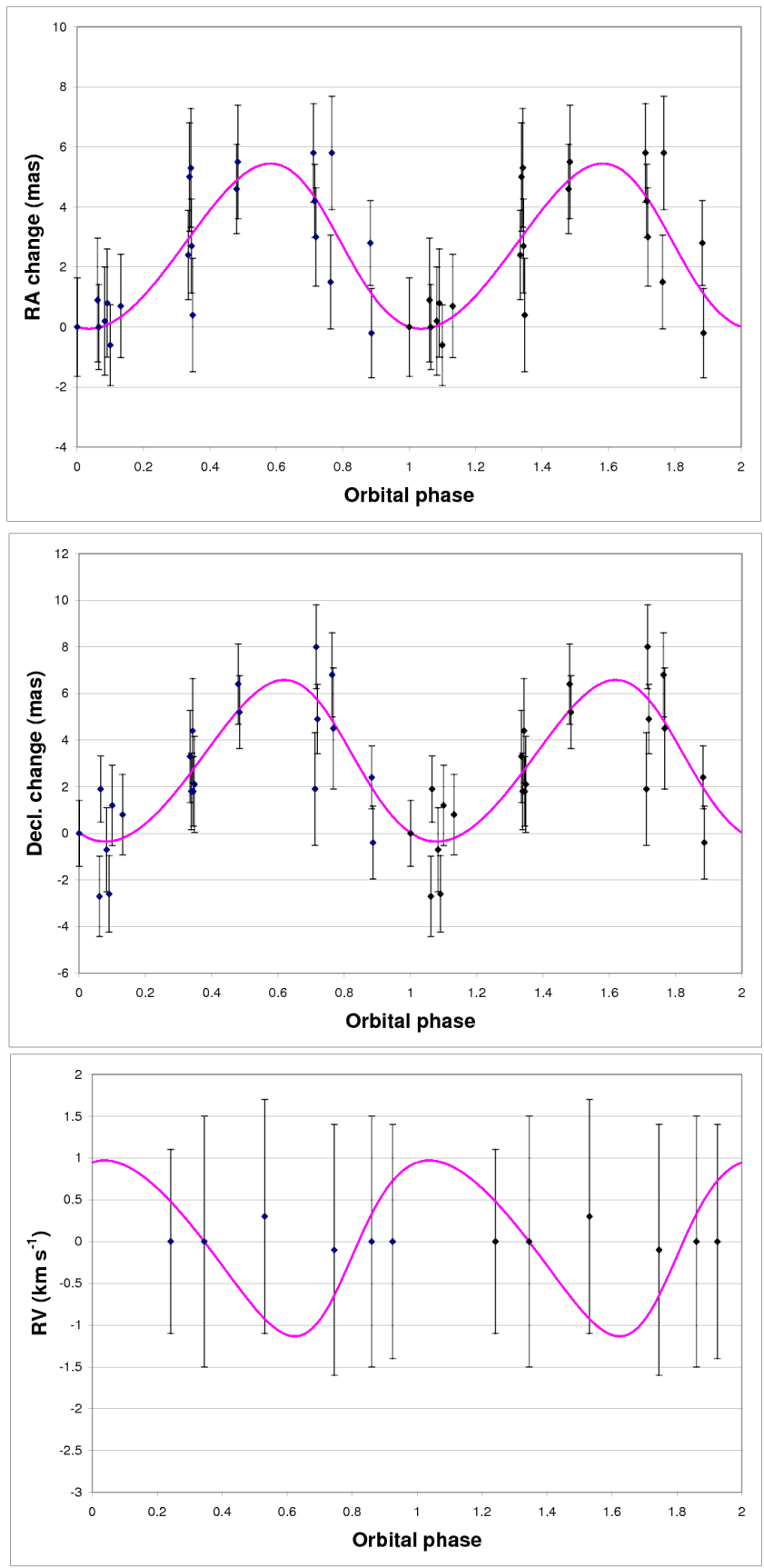

**Figure 6**

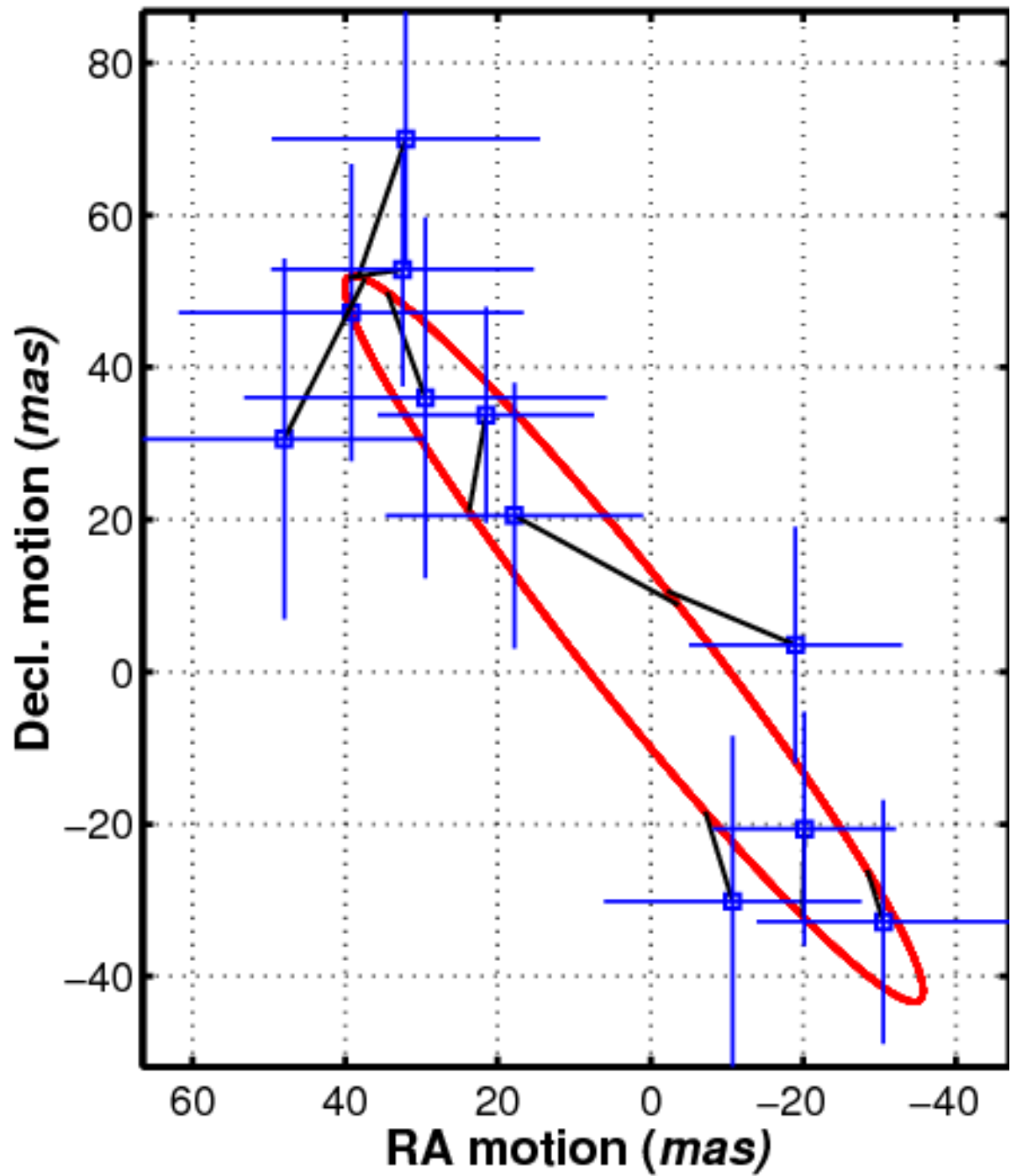

**Figure 7**

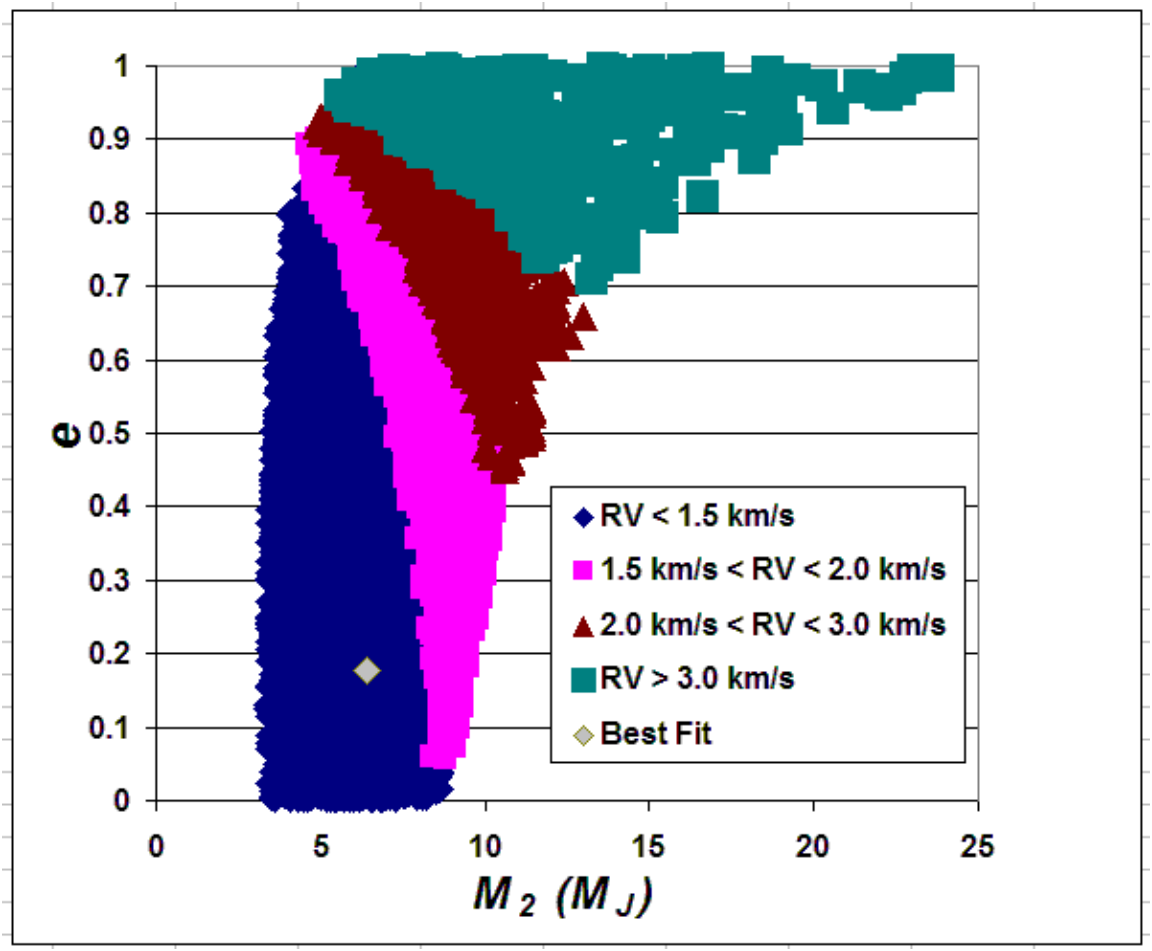

**Figure 8**

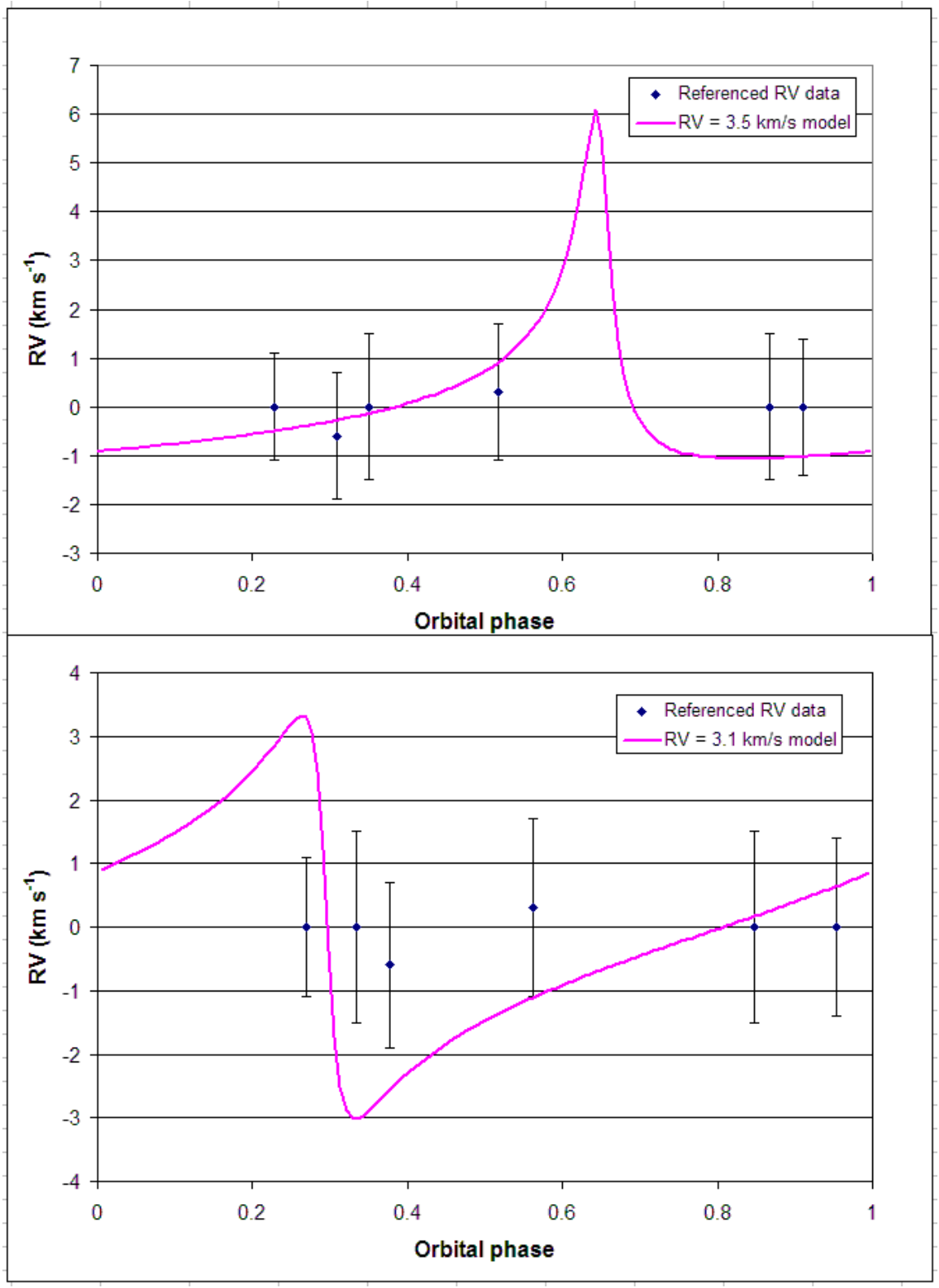

**Figure 9**